\documentclass[12pt,tightenlines,showpacs,preprintnumbers,superscriptaddress,nofootinbib]{revtex4-1}
\pdfoutput=1

\usepackage[pdftex]{graphicx}
\usepackage[T1]{fontenc}
\usepackage{lmodern}
\usepackage{verbatim}

\usepackage{epsfig}

\usepackage{bm}        
\usepackage{amssymb}  
\usepackage{mathrsfs}   
\usepackage{cancel}      
\usepackage{bm}
\usepackage{braket}
\usepackage{amsmath}
\usepackage{color}
\usepackage{multirow}

\newcommand{\be}{\begin{equation}}
\newcommand{\ee}{\end{equation}}
\newcommand{\bea}{\begin{eqnarray}}
\newcommand{\eea}{\end{eqnarray}}
\newcommand{\df}{\dfrac}

\newcommand{\PRE}[1]{{#1}} 

\begin{document}

\preprint{ULB-TH/14-22}

\title{ \PRE{\vspace*{1.5in}}  SIMPLIFIED SIMPs AND THE LHC
\PRE{\vspace*{0.3in}} }

\author{N. Daci}
\author{I. De Bruyn}
\author{S. Lowette}

\affiliation{Vrije Universiteit Brussel\\
Inter-University Institute for High Energies\\
Pleinlaan 2, 1050 Brussels, Belgium\PRE{\vspace*{.2in}}}

\author{M.H.G. Tytgat\PRE{\vspace*{.2in}}}
\author{B. Zaldivar\PRE{\vspace*{.2in}}}

\affiliation{Service de Physique Th\'eorique\\
 Universit\'e Libre de Bruxelles\\ 
Boulevard du Triomphe, CP225, 1050 Brussels, Belgium\PRE{\vspace*{.2in}}}



\begin{abstract}
The existence of Dark Matter (DM) in the form of Strongly Interacting Massive Particles (SIMPs) may be motivated by  astrophysical observations that challenge the classical Cold DM scenario. Other observations greatly constrain, but do not completely exclude, the SIMP alternative. 
The signature of SIMPs at the LHC may consist of neutral, hadron-like, trackless jets produced in pairs.
We show  that the absence of charged content can provide a very efficient tool to suppress dijet backgrounds at the LHC, thus enhancing the sensitivity to a potential SIMP signal. We illustrate this using a simplified SIMP model and present a detailed feasibility study based on simulations, including a dedicated detector response parametrization. We evaluate the expected sensitivity to various signal scenarios and tentatively consider the exclusion limits on the SIMP elastic cross section with nucleons.
\end{abstract}

\maketitle

\newpage

\section{ \label{sec:intro} Introduction}

There is ample evidence of the existence of Dark Matter (DM) in the Universe. First invoked to explain a puzzle in clusters of galaxies \cite{Zwicky:1933gu}, the case for DM, based on observations from galactic up to the largest scales, is now very strong (see for instance \cite{Peebles:2013hla,Bertone:2004pz}). In particular, precision measurements of the cosmic microwave background anisotropies imply that about 80\% of the matter content of the Universe is made of DM \cite{Komatsu:2010fb,Ade:2013zuv}. Despite this, the  precise nature of DM remains a mystery. The most-studied hypothesis assumes that DM is made of weakly interacting massive particles or WIMPs. This rests on the observation that a stable massive particle with annihilation cross section of order $1$~pb --characteristic of weak interactions-- could have a relic abundance that agrees with cosmological measurements. A popular WIMP is the neutralino, the typical DM candidate of supersymmetric extensions of the Standard Model (SM) of particle physics \cite{Jungman:1995df}, but many alternative WIMP candidates have been proposed \cite{Bergstrom:2009ib}, which are actively being searched for by experiments, including those at the LHC. Yet, as appealing as the WIMP scenario may be, it remains important to study other possibilities. Amazingly enough, a strongly interacting massive particle (SIMP), {\em i.e.} a particle with strong interactions with ordinary baryons, is not yet fully excluded \cite{Starkman:1990nj} (see also \cite{McGuire:2001qj} and specially \cite{Mack:2007xj} for a more recent appraisal of existing constraints). While a SIMP scenario may seem exotic at first sight, it may be motivated by the long lasting interest for DM particles with strong self-interactions, going back to the seminal work of Spergel and Steinhardt  \cite{Spergel:1999mh}. Indeed, self-interacting dark matter (SIDM) particles\footnote{Unfortunately, strongly interacting and self-interacting share the same acronym, SI, so that in the literature SIDM may stand for strongly interacting dark matter or SIMP for self-interacting massive particles (see e.g.\cite{Hochberg:2014dra}).} with very large cross section, $\sigma_{\chi\chi}/m_\chi \sim 10^{-24}$ cm$^2/$GeV, may help addressing astrophysical observations that present a challenge for the cold dark matter (CDM) paradigm, like the missing satellites or core-cusp problems \cite{Bullock:2010uy,BoylanKolchin:2011de,Weinberg:2013aya,Famaey:2013ty}. While it is possible to build scenarios with a strongly interacting hidden sector weakly coupled to the SM particles (see {\em e.g.} \cite{Han:2007ae}), it is perhaps as natural to consider SIDM particles with strong interactions with ordinary matter. The latter possibility is  much more constrained but, again, it is not fully excluded \cite{Starkman:1990nj}. This will be our main motivation to take the SIMP hypothesis seriously. In any event, the absence of a clear signal in recent WIMP searches may be a further motivation to look away from the lamppost and to explore a bit more this exotic possibility. Here we focus in particular on the possible search of a SIMP candidate at the LHC.

In particular we consider the possibility of observing events with no (or little) signal in the tracking systems and electromagnetic calorimeters and only (or essentially) energy deposition in the hadron calorimeters of the detectors, akin to the signature of neutrons or $K_L^0$. As far as we know, this possibility has been first put forward in \cite{Mohapatra:1999gg}, albeit in the framework of the Tevatron, and  more recently in \cite{Bai:2011wy}. We will follow the simplified model approach of the latter work but our feasibility study goes quite a few steps further. In particular, we develop the charged contents as a discriminator to suppress dijet backgrounds at LHC, thus enhancing the sensitivity to a potential SIMP signal. We present a feasibility study based on simulations, including a \textsc{Delphes}~\cite{deFavereau:2013fsa} description of a typical LHC detector, and evaluate the expected sensitivity to various signal scenarios. We finally tentatively map the expected sensitivity onto the elastic cross section for scattering of SIMPs on nucleons and compare the resulting exclusion limit (i.e. assuming no signal is seen at the LHC) to other constraints, mostly astrophysical but also from direct detection DM searches. 

The plan of this article is as follows. Firstly, we briefly motivate the models we consider in the light of the existing constraints (Section~\ref{sec:mod}). Next we describe the expected signature at the LHC, and the dedicated simulation of the signal and background samples (Sections~\ref{sec:generation} and~\ref{sec:simulation}), and detail the strategy of our analysis (Section~\ref{sec:strat}), before giving the resulting expected sensitivity on various signal scenarios (Section~\ref{sec:res}). In the last section, we tentatively re-express this  in terms of an exclusion limit on the SIMP-nucleon cross section (Section~\ref{sec:comp}) and finally draw our conclusions (Section~\ref{sec:con}). 

Before to go on, we  mention that other scenarios with a strongly coupled hidden sector that may lead to LHC signals somewhat analogous to trackless jets have been proposed  recently, see~\cite{Schwaller:2015gea,Cohen:2015toa}. If necessary, those may provide a further motivation for the kind of experimental study we discuss in the following sections.

\section{ \label{sec:mod} SIMPLIFIED SIMP MODELS}

In this section we motivate the SIMP models that we consider. Taking into account the existing constraints, the possibilities boil down to a few  options \cite{Bai:2011wy}, at least provided we focus on simple models. By this we mean assuming that the SIMP, which may be composite, can nevertheless be treated as an elementary particle on all energy scales we deal with. We will furthermore suppose that the SIMP particles interact with SM particles (here quarks) through a mediator elementary particle. In short, we apply the philosophy of simplified models used for WIMP searches at colliders \cite{Abdallah:2014hon,Malik:2014ggr} to SIMP phenomenology \cite{Bai:2011wy}. Given the exploratory character of the search for  DM it may be reasonable to begin with such simple assumptions. Whether the conclusions that can be drawn using such a framework are generic is another question. The dynamics of strong interactions of the ordinary hadrons is very complex. The same is probably true of realistic SIMP scenarios, see {\em e.g.} \cite{Farrar:1984gk}. Yet, the phenomenological approach we advocate allows to go significantly beyond the approximation  usually adopted to describe SIMP interactions. In particular, it allows to confront SIMP properties advocated to solve astrophysical issues, with possible signatures at high energies. In the rest of this section, we summarize the properties of our simplified SIMP models, following essentially the arguments of \cite{Bai:2011wy}.

One of the strongest constraints on a SIMP as DM is set by searches for heavy isotopes, in particular heavy water, which put limits on the formation of bound states between SIMPs and nucleons. Assuming that the SIMP is the dominant form of DM, a particle lighter than $\sim 10 $ TeV that can form a bound is excluded \cite{Starkman:1990nj} (see also \cite{Burdin:2014xma}).
This constraint is  avoided if the SIMP has purely repulsive SIMP-nucleon interactions, which may be achieved if it interacts with SM particles through a scalar or vector mediator with opposite sign couplings. In the case of a vector mediator we should be concerned with the fact that vector mediators couple to DM antiparticles  with an opposite charge \cite{Bai:2011wy}. This is avoided if there are no DM antiparticles around, that is if the abundance of DM is asymmetric. This could actually be the case. First, a symmetric SIMP candidate can only be a sub-dominant component of DM  if its abundance is set by thermal freeze-out. Conversely, if most of DM is made of SIMPs, then its abundance is determined either by an asymmetry or through a non-thermal mechanism. Second, measurements of the Earth heat flow set strong constraints on SIMP properties. For cross sections that are characteristic of a SIMP, DM could be efficiently captured and accreted in the core of the Earth. Annihilating SIMPs would provide a substantial source of heat, a constraint that does not apply to asymmetric candidates \cite{Mack:2007xj}.\footnote{Notice however that, in an asymmetric scenario, light scalar DM particles can lead to black hole formation if they are trapped inside neutron stars \cite{Kouvaris:2011fi,McDermott:2011jp}. This is valid for light asymmetric bosonic DM candidates. For simplicity, we only consider fermionic DM candidates.} In the sequel we will  take seriously the possibility that DM is made dominantly of a SIMP, if anything because it may be of interest to compare the LHC reach with the expectations from other DM searches, like direct detection ones for example, which implicitly rely on this assumption.    


The interaction Lagrangian of the models we will consider is then simply (see also \cite{Bai:2011wy}) 
\be
\label{eq:model}
{\cal L}_{int} = \left\{ 
\begin{array}{lll}
- g_\chi \phi \,\bar{\chi} \chi - g_q \phi \,\bar{q} q& & \mbox{(scalar mediator)}\\
\\
- \tilde g_\chi  \phi_\mu \,\bar{\chi}\gamma^\mu \chi - \tilde g_q \phi_\mu \,\bar{q}\gamma^\mu q & & \mbox{(vector mediator)}
\end{array}\right.
\ee
with $g_\chi g_q, \tilde g_\chi \tilde g_q<0$ to avoid the  formation of bound states. Further parameters of the model are the mass of the messenger, $m_\phi$, and of the SIMP, $m_\chi$. The model has thus at least 4 free parameters. For LHC phenomenology, only the product of the two couplings appears, but astrophysics constrains both DM self-interaction and interactions with ordinary matter. Furthermore, although we have in mind the fact that SIMPs should\footnote{Light SIMPs significantly coupled to b or c quarks is probably  constrained by B and D meson phenonomenology.} have flavour dependent couplings, for simplicity we assume in this study that they have a universal coupling to quarks.  

Introducing new strong interactions between quarks, and thus nucleons, is  not harmless. In the sequel we will consider a rather light mediator, $m_\phi \sim 1$ GeV, as in  \cite{Bai:2011wy}. While the precise value of this mass may seem {\em a priori}  of little importance for SIMP production at the LHC, we will argue that it is crucial to assess  their detection in the hadron calorimeters as well as for the comparison with other DM searches. In order to keep small the impact of the new interaction on the nuclear potential, we will furthermore assume, again as in \cite{Bai:2011wy}, that the mediators do not modify nuclear potentials by more than ${\cal O}(10 \%)$, so that  $g_{\chi N} \lesssim 0.3 g_{\pi N} \sim 3$   for  $m_\phi \sim m_\pi$, where $g_{\pi N} \sim 13$ is the effective pseudoscalar pion-nucleon coupling \cite{Donoghue:1992dd}, 
or, for  $m_\phi\sim 1$ GeV, $\tilde g_{\chi N} \lesssim 0.3 g_{\rho N} \sim 6$, where $g_{\rho N}\sim18$ \cite{Tuchitani:2004ur} is the vector $\rho$ meson coupling to the nucleon. This will be relevant in Section~\ref{sec:comp} when we will discuss the comparison between high energy and low energy constraints. 

There are further constraints on the interaction between DM and ordinary matter, and between DM particles themselves that we should take into account. 
First, one may look for elastic scattering of SIMPs with nuclei in direct detection experiments. However, as a SIMP interacts strongly in the Earth (or even in the atmosphere), it cannot reach the underground direct detection detectors \cite{Albuquerque:2003ei}.\footnote{Some interesting implications for direct detection experiments of significant energy loss of DM in the Earth are discussed in \cite{Kouvaris:2014lpa}.} Instead, such high interaction cross sections are constrained by space or airborne experiments like RSS, a balloon-based detector \cite{Rich:1987st} and XQC, a sounding rocket X-ray experiment \cite{Erickcek:2007jv}. There are also constraints  from primordial nucleosynthesis and cosmic rays \cite{Cyburt:2002uw,Mack:2013ofa}. All these constraints have been extensively reviewed in \cite{Mack:2007xj}, to which we refer for more details since we have nothing specifically new to say about them. Finally, there are strong constraints on interactions between DM and baryons from observations of  Cosmic Microwave Background Radiation (CMBR) anisotropies and  large structure, including from Lyman-$\alpha$ data \cite{Chen:2002yh,Dvorkin:2013cea}. In particular the constraints reported in \cite{Dvorkin:2013cea} are relatively new and somewhat stronger than previously thought. We will discuss them further in section IV where we also discuss the cosmological constraints on DM (self)interactions. 


The possible signatures of SIMPs at colliders have been much less studied \cite{Mohapatra:1999gg,Bai:2011wy}. The production at LHC of WIMPs is currently studied through missing transverse energy signals, typically jets or photons with large missing momentum~\cite{Goodman:2010ku,Goodman:2010yf,Bai:2010hh,Khachatryan:2014rra,Aad:2015zva}. Such constraints also apply to SIMPs, provided their interaction with baryons is less than that characteristic of hadrons, so that the DM particles produced do not deposit substantial energy in the calorimeters of the detectors. In the present work, we discuss further the complementary possibility of observing trackless jets from DM interactions.

\section{ \label{sec:sign} SIMP observability at the LHC}

The potentially high interaction cross section of the SIMPs with nuclear matter leads to a particular signature at the LHC. Indeed, the neutral SIMP leaves no track in a tracking detector, but will give rise to a highly-energetic hadronic shower. Since a pair of SIMPs will be dominantly produced back-to-back in the transverse plane, this leads to the peculiar observable signature of a dijet pair without any tracks. Here, we study the feasibility of detecting this signature at the LHC, extending on the previous work in~\cite{Bai:2011wy}. 

\subsection{Event generation}
\label{sec:generation}

The interaction Lagrangian (\ref{eq:model}) was implemented in \textsc{FeynRules 2.0}~\cite{Alloul:2013bka} for the scalar mediator case and subsequently interfaced to \textsc{Madgraph 5}~\cite{Alwall:2014hca} to generate the pair production of SIMPs in proton--proton collisions. The center-of-mass energy was chosen to be $\sqrt{s} = 8\,\mathrm{TeV}$, corresponding to the energy at which the LHC delivered collisions in 2012. The benchmark for our simulation is defined by the couplings $g_\chi = -1$, $g_q = 1$ and the mediator mass to $m_\phi = 1$ GeV (the precise value of $m_\phi$  plays no role at this level of the discussion). 
We consider a stable dark-matter particle $\chi$, and generated events for various masses $m_\chi = 1$, $10$, $100$, $200$, $400$, $700$, and $1000 \,\text{GeV}$.

In Table~\ref{tab:xsec}, the SIMP production cross section is shown for each considered SIMP mass $m_\chi$, along with the number of events expected for an integrated luminosity of $20 \, \mathrm{fb}^{-1}$, which more or less corresponds to the dataset recorded by the LHC experiments in 2012. These values were obtained using $|\eta(\chi)| < 2.5$ and $p_{\rm T}(\chi) > 250 \, \mathrm{GeV}$ pres-election requirements.
Along with the case of a scalar mediator, the consistently larger values for a vector mediator are reported as well.
For this LHC feasibility study, we have focused our simulations on the scalar case, but the results can be directly translated to a vector mediator by straightforward scaling of the reported cross sections.

\begin{table}[h]
       \centering
\renewcommand{\tabcolsep}{.2cm}
	\renewcommand{\arraystretch}{1.2}
	\begin{tabular}{| c | l | r | l | r |}
	\hline
	 & \multicolumn{2}{c|}{Scalar} & \multicolumn{2}{c|}{Vector} \\
	\hline
	$m_\chi$ [GeV] & $\sigma_{\bar{\chi}\chi}$ [pb] & Events/$20 \text{fb}^{-1}$ & $\sigma_{\bar{\chi}\chi}$ [pb] & Events/$20 \text{fb}^{-1}$ \\
	\hline
	   1 &   2.80     &  56040 &  3.18    &  63645 \\
	  10 &   2.79     &  55800 &  3.17    &  63400 \\
	 100 &   1.88     &  37620 &  2.49    &  49826 \\
	 200 &   0.728    &  14558 &  1.31    &  26196 \\
	 400 &   0.0769   &   1539 &  0.229   &   4583 \\
	 700 &   0.00363  &     73 &  0.0167  &    336 \\
	1000 &   0.000239 &      5 &  0.00147 &     31 \\
	\hline
	\end{tabular}
	\caption{\label{tab:xsec} 
		Production cross section for each SIMP mass, and number of events corresponding to an integrated luminosity of $20\,\mathrm{fb}^{-1}$, after $|\eta(\chi)| < 2.5$ and $p_{\rm T}(\chi) > 250 \, \mathrm{GeV}$ pres-election requirements.}
\end{table}

At these high cross sections of dijet-like signal events, we consider QCD jet production as the main background process to take into account. Also this background was generated with the \textsc{MadGraph} program, where we produced two samples in the $] 500 , 1000 ]$ and $] 1000 , \inf [ \,\mathrm{GeV}$ bins of $H_{\rm T} = \sum_{\rm partons} \left| p_{\mathrm{T},i} \right|$, the scalar sum of the transverse momenta of the outgoing partons. The corresponding cross sections are respectively $8426$ and $204 \,\mathrm{pb}$. With the above settings, the possible additional production of dijet events through the mediator $\phi$ was verified to be at the percent level with respect to the QCD dijet production, and is further ignored.

Events from both signal and background samples are subsequently processed with \textsc{Pythia8}~\cite{Sjostrand:2006za}, using the \texttt{CTEQ6L1}~\cite{Pumplin:2002vw} parton distribution functions, in order to embed the hard interactions in full proton collisions, including the description of the parton shower, hadronization, and underlying event.

\subsection{Detector simulation}
\label{sec:simulation}


In order to study the observability of a SIMP signal at the LHC, we use the parametrized detector simulation package \textsc{Delphes}~\cite{deFavereau:2013fsa}. Using this program, we simulate the response of a typical LHC detector to the generated signal and background collisions. We use \textsc{Delphes} in the standard CMS~\cite{Chatrchyan:2008aa} configuration, which implements spatial and energy resolution functions for each sub-detector --- inner tracker, electromagnetic and hadronic calorimeter, and muon spectrometer --- and adds high-level event reconstruction, like jet clustering using \textsc{FastJet}~\cite{Cacciari:2011ma}. Some parts of this event reconstruction have particular relevance for this analysis.

First, the track reconstruction plays an important role in establishing jets to arise from only neutral particles. In \textsc{Delphes}, tracks are built from generated particles, to which a realistic inefficiency function is applied. Hence, only genuine tracks are being simulated. This is adequate for our purpose, since we expect our QCD dijet background to arise from a combination of tracking inefficiency and of fluctuations in the jet fragmentation, leading to a genuinely small charged jet content.

Further, an adequate simulation of the jet response is needed. This is obtained in \textsc{Delphes} through the smearing of jet energy measurements with realistic jet resolution functions, rather than using a detailed simulation of the calorimeter response. For the SIMPs, though, we implemented a more elaborate approach. The interaction of the SIMP $\chi$ with the detector's calorimeters can be described as an interaction of a hadron with the nuclei of the detector material, but comes with a potentially different nucleon interaction cross section $\sigma_{\chi N}$. We model this interaction in a way that allows to easily simulate a change of $\sigma_{\chi N}$, for instance leading to incomplete containment of the deposited energy in the calorimeters.

The position of the first $\chi N$ collision serves as the starting point of an ensuing hadronic shower. This initial position is distributed exponentially as $e^{-\ln(2) \frac{\sigma_{\chi N}}{\sigma_{\rm QCD}}\frac{x}{\lambda_I}}$ , where $x$ is the calorimeter depth in interaction lengths $\lambda_I$ and $\sigma_{\rm QCD}$ is the standard hadronic interaction cross section of $\sim 40 \, \mathrm{mb}$, which is taken constant in the energy range of the collisions of interest. We study the dependence of the search sensitivity on $\sigma_{\chi N}$ by considering $\sigma_{\chi N} / \sigma_{\rm QCD} = 1$, where full shower containment is expected, and $\sigma_{\chi N} / \sigma_{\rm QCD} = 0.1$, where late shower development will lead to calorimeter energy leakage. As will be detailed later, this lower value allows to gauge the dependence of the analysis on $\sigma_{\chi N}$. A larger value of $\sigma_{\chi N}$ was not explicitly studied, but can up to some level still be assumed to lead to full registration of the energy of the induced early showers.

After the first collision, the shower will develop as a mixture of subsequent collisions of the SIMP with standard hadronic collisions of particles created in the shower. For simplicity, we assume standard longitudinal hadronic shower development. This is expected to be a good approximation, since the energy loss of the SIMP is largest in the beginning of the shower. We model the longitudinal energy profile of the shower according to~\cite{Bock:1980rs}, with parameters estimated for iron absorber~\cite{Hughes:1990hm,Kulchitsky:1998wj}.
The total deposit of energy in the calorimeters is then calculated as an integration of the longitudinal energy profile, from the starting position of the shower up to the rear face of the calorimeter. For this, we assume a calorimeter depth of $1.7 \lambda_I$ and $9 \lambda_I$ for our electromagnetic and hadronic calorimeters respectively, uniform as a function of polar angle of the incident particle.

Finally, due to the high instantaneous luminosities reached by the LHC, additional proton--proton interactions, called \emph{pile-up}, may occur simultaneously in each LHC bunch crossing. These pile-up collisions are also simulated by \textsc{Delphes}, where we set the average number of pile-up interactions to 21, Poisson distributed, and with a resolution on the $z$ coordinate of the interaction vertices $\sigma_z = 0.1 \,\mathrm{mm}$.
This configuration approximates sufficiently the reality of the data-taking in the 2012 LHC run.

For events containing pile-up interactions, we apply the \textsc{Delphes} pile-up subtraction algorithm in order to remove the additional particles and their energy deposits from the reconstruction. This is, on the one hand, important to avoid charged particles from the pileup interactions that overlap with SIMP jets to induce inefficiencies in the signal reconstruction. It was checked that this does not happen at an appreciable level. On the other hand, the energy deposits from pile-up collisions may bias jet energy estimates, affecting the analysis selection. Also this effect was checked, and the pile-up subtraction was found to remove any such significant pile-up dependence.

\subsection{Analysis}
\label{sec:strat}

The SIMP signal considered in this paper would manifest itself in the detector as a pair of trackless jets, back-to-back in the plane transverse to the beamline. The background to such a signal is expected to be dominated by standard QCD dijet production, where an interplay of rare jet fragmentation and tracking inefficiency leads to a very low number of tracks in the jet, and a very low fraction of jet energy emitted in charged particles. To quantify the latter, we use as an observable the so-called charged energy fraction (CHEF), defined as the ratio $\sum_i p_{\rm T,i} / p_{\rm T,jet}$, where the sum runs over all tracks with transverse momenta $p_{\rm T,i}$, associated to the jet with transverse momentum $p_{\rm T,jet}$.

\begin{figure}
\centering
  \centering
  \includegraphics[width=0.48\linewidth]{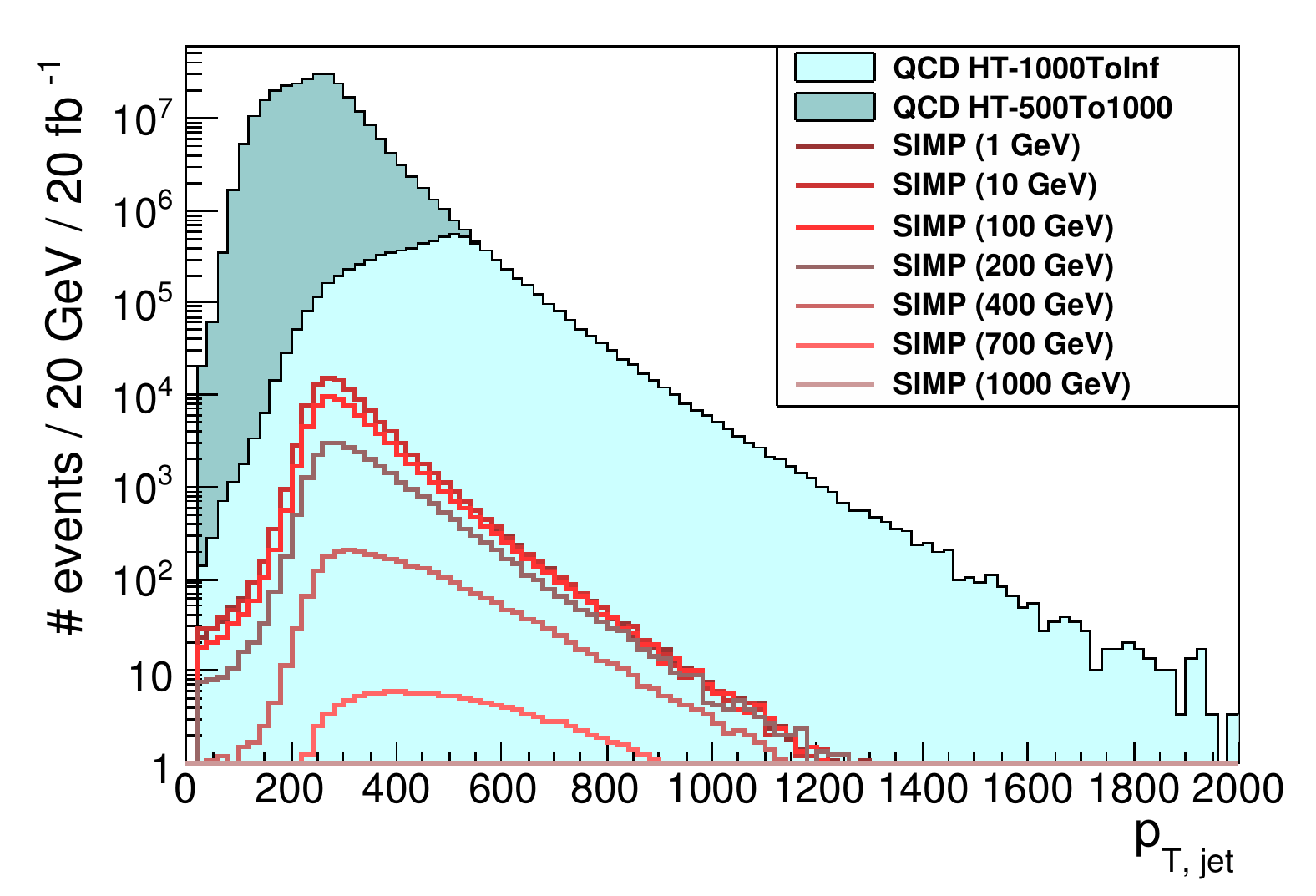}
  \hfill
  \includegraphics[width=0.48\linewidth]{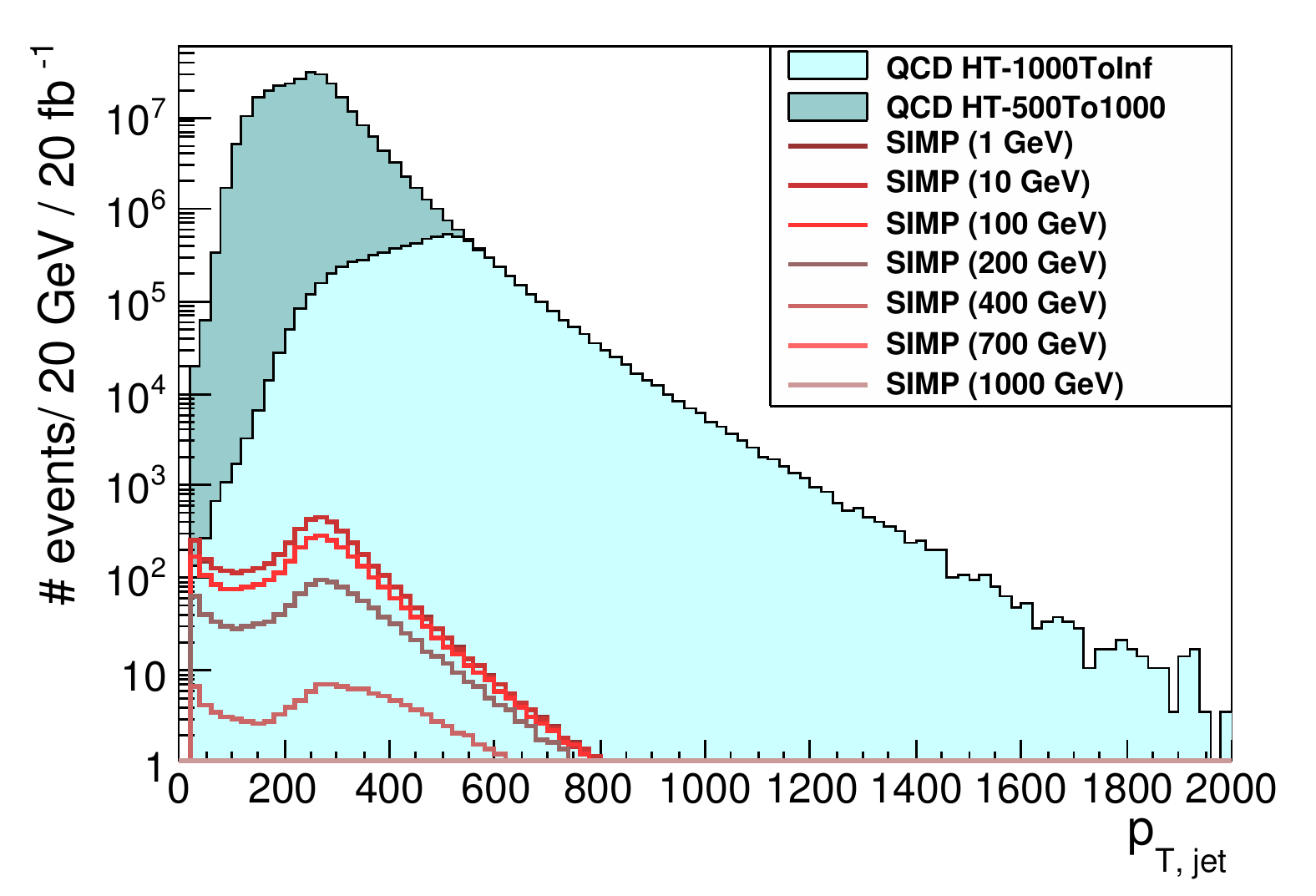}
\caption{The jet $p_T$ of the two leading jets after the selection cuts described in the text, excluding the jet $p_T$ cut, for $\sigma_{\chi N} / \sigma_{\rm QCD} = 1$ (left) and $\sigma_{\chi N} / \sigma_{\rm QCD} = 0.1$ (right). In all cases, a generator-level $p_T$ pres-election at 250GeV is already applied.}
\label{fig:jet2pt}
\end{figure}

The selection of the signal events with high-$p_{\rm T}$ back-to-back jets proceeds as follows. At least two jets are required with $p_{\rm T} > 350 \, \mathrm{GeV}$ and $| \eta_\text{jet} | < 2.0$, with an azimuthal separation $ | \Delta \phi_\text{jj} | > 2.0 $ on the two leading jets. The $p_{\rm T}$ requirement on the jets is driven by a typical dijet trigger requirement for 2012 LHC conditions, while the $\eta$ restriction ensures the tracks of the jets to be well contained within the tracking detector. In Figure~\ref{fig:jet2pt}, the $p_{\rm T}$ of both leading jets is shown after the above cuts, excluding the $p_T$ cut, for all considered signals and for the background, for both cases $\sigma_{\chi N} / \sigma_{\rm QCD} = 1$ and $\sigma_{\chi N} / \sigma_{\rm QCD} = 0.1$. The similarity of the shape of signal and background distributions at high $p_{\rm T}$ is a manifestation of the lack of possible discrimination of signal and background on a purely kinematic basis.

\begin{figure}
\centering
  \centering
  \includegraphics[width=0.49\linewidth]{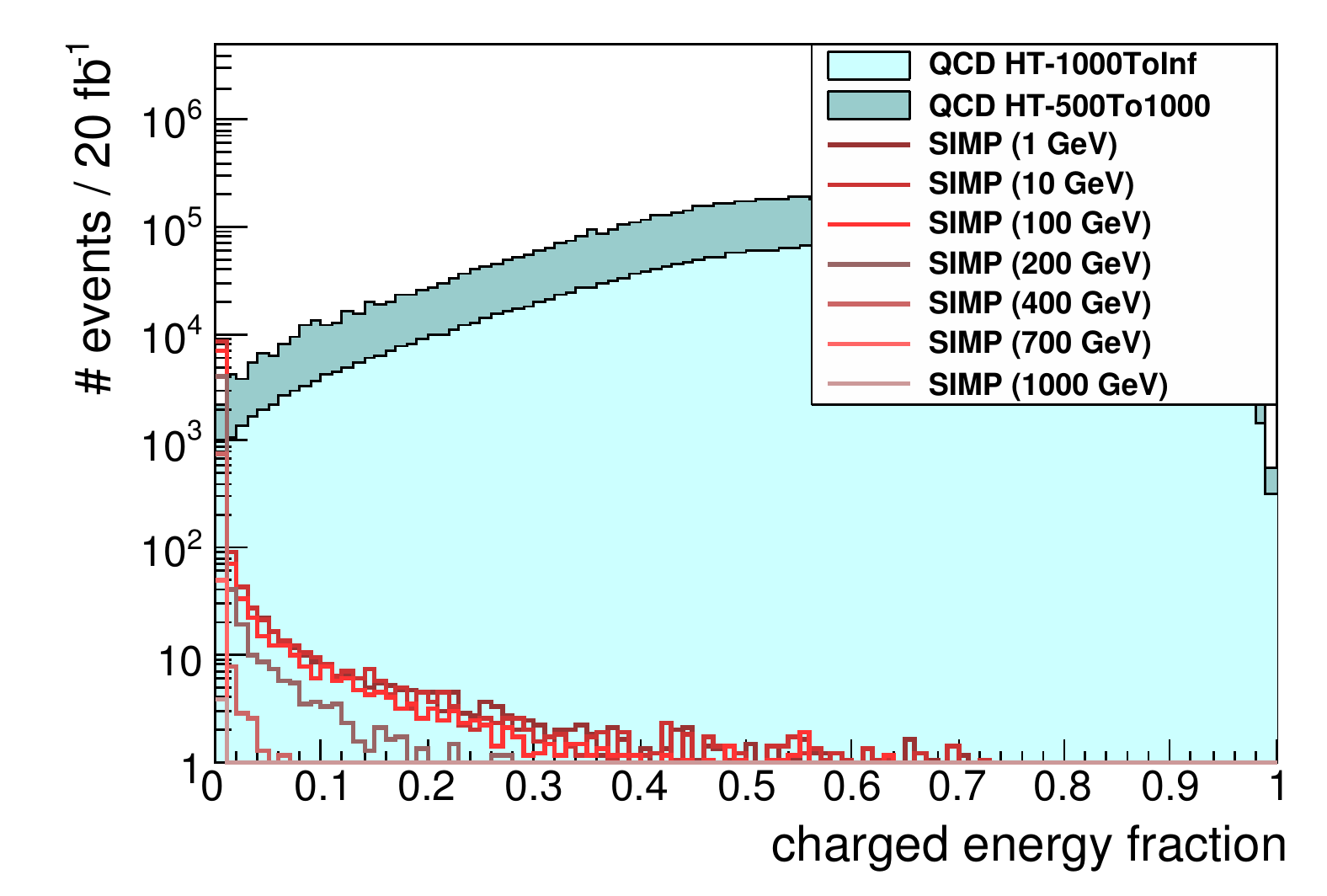}
  \hfill
  \includegraphics[width=0.48\linewidth]{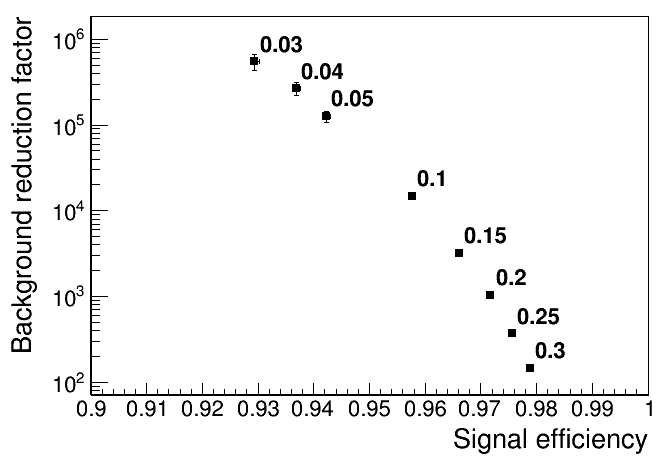}
\caption{Left: charged energy fraction of the highest-$p_{\rm T}$ jet, after kinematic selection requirements as described in the text. Right: Background suppression factor versus signal selection efficiency for various cuts on the CHEF, labeled on the graph.}
\label{fig:chef}
\end{figure}

The number of tracks or the CHEF of a jet provide strong handles to suppress the background to the SIMP signal. In the remainder, the CHEF is used, as it was found to have a better performance than the number of tracks. The distribution of the CHEF observable is shown in Figure~\ref{fig:chef} (left) for the leading jet, comparing signal to background, after the kinematic cuts above. In Figure~\ref{fig:chef} (right), the background suppression is compared to the signal selection efficiency, for various cuts on the CHEF of both the two highest-$p_{\rm T}$ jets. A background rejection with a factor of 10000 can be achieved with a signal selection efficiency above 95\%.

While the CHEF is the main handle to separate a SIMP signature from QCD jets, other differences between signal and background exist. These arise due to the nature of SIMP showers being different than the QCD background jets, the latter arising from quark or gluon fragmentation. The SIMP, being a single particle, is expected to induce a narrower hadronic shower, potentially with a different longitudinal shower development, leading to differences in the shower shape, the jets' electromagnetic fraction, or giving rise to increased punch through of particles from the hadronic showers in the muon system. Although exploiting such effects can significantly enhance further the analysis sensitivity~\cite{Aad:2015asa}, it requires more detailed simulations, beyond the scope of this study.

The cut on the CHEF of the two leading jets is chosen to ensure a large signal-to-background ratio over a wide range of SIMP masses. Contrary to many searches at the LHC, not much analysis optimisation is possible, nor needed. On the kinematic side, the cuts are fully defined by trigger and detector acceptance, as described above. What concerns the CHEF cut, it is mostly defined by the available integrated luminosity, since the production cross section of the SIMPs cannot vary much without losing the signal, despite a large signal-to-background cross section ratio. Indeed, as the SIMP inelastic interaction and related production cross sections decrease, the signal will quickly transition from a visible pair of trackless jets into missing energy, and the analysis presented in this paper unavoidably loses its sensitivity. In order to provide a way to gauge this effect, we assessed the search sensitivity on the samples simulated with a factor 10 reduction in the inelastic interaction cross section $\sigma_{\chi N}$, which is proportional to $g_\chi^2 g_q^2$. Because of the same proportionality, this leads to an equal reduction by a factor of ten of the production cross sections mentioned in Table~\ref{tab:xsec}. This latter reduction is further always implied when considering $\sigma_{\chi N} / \sigma_{\rm QCD} = 0.1$.

\subsection{Results}
\label{sec:res}

In Table~\ref{tab:results_1}, we summarize the number of events after the various analysis cuts, for all signal and background samples, assuming $\sigma_{\chi N} / \sigma_{\rm QCD} = 1$. Table~\ref{tab:results_0.1} contains the results for $\sigma_{\chi N} / \sigma_{\rm QCD} = 0.1$. To overcome the statistical limitation from the finite simulated background samples, we calculated the number of QCD multijet background events passing all cuts by assuming uncorrelated efficiencies of the CHEF requirements on the two leading jets. We checked this assumption to be conservative, in the sense that for loose cuts on the jets' CHEF, the assumption of uncorrelated cuts leads to a similar or smaller background reduction compared to a direct estimation from applying the CHEF requirement on both jets in the simulated events.

\begin{table}[ht]
\centering
\renewcommand{\arraystretch}{1.2}
\renewcommand{\tabcolsep}{.3cm}
\begin{tabular}{| c | c | c | c | c | c | c |}
 \hline
  \multirow{2}{*}{Sample} & \multirow{2}{*}{Total} & \multirow{2}{*}{Selection} & \multicolumn{4}{c|}{Charged Energy Fraction} \\
  \cline{4-7}
   & & & 0.1 & 0.05 & 0.04 & 0.03 \\
   \hline
   QCD HT-500To1000 & 168520000 & 4992040 & 378 & 45 & 21 & 11 \\
   QCD HT-1000ToInf & 4080000 & 2691870 & 144 & 16 & 8 & 3 \\
   SIMP (1 GeV) & 56040 & 9076 & 8693 & 8553 & 8504 & 8436 \\
   SIMP (10 GeV) & 55800 & 9060 & 8674 & 8531 & 8479 & 8413 \\
   SIMP (100 GeV) & 37620 & 7252 & 6950 & 6841 & 6804 & 6748 \\
   SIMP (200 GeV) & 14558 & 4072 & 3928 & 3871 & 3848 & 3821 \\
   SIMP (400 GeV) & 1539 & 771 & 749 & 739 & 735 & 730 \\
   SIMP (700 GeV) & 73 & 52 & 51 & 50 & 50 & 49 \\
   SIMP (1000 GeV) & 5 & 4 & 4 & 4 & 4 & 4 \\
   \hline
 \end{tabular}
 \caption{\label{tab:results_1} 
	  The number of events after the selection cuts and several cuts on the charged energy fraction, for the 2 background samples and the 7 different signal samples with $\sigma_{\chi N} / \sigma_{\rm QCD} = 1$.}
\end{table}

\begin{table}[ht]
\centering
\renewcommand{\arraystretch}{1.2}
\renewcommand{\tabcolsep}{.3cm}
\begin{tabular}{| c | c | c | c | c | c | c |}
 \hline
  \multirow{2}{*}{Sample} & \multirow{2}{*}{Total} & \multirow{2}{*}{Selection} & \multicolumn{4}{c|}{Charged Energy Fraction} \\
  \cline{4-7}
   & & & 0.1 & 0.05 & 0.04 & 0.03 \\
   \hline
   QCD HT-500To1000 & 168520000 & 4992040 & 378 & 45 & 21 & 11 \\
   QCD HT-1000ToInf & 4080000 & 2691870 & 144 & 16 & 8 & 3 \\
   SIMP (1 GeV) & 5604 & 146 & 134 & 131 & 131 & 130 \\
   SIMP (10 GeV) & 5580 & 144 & 132 & 130 & 129 & 128 \\
   SIMP (100 GeV) & 3762 & 117 & 107 & 105 & 105 & 104 \\
   SIMP (200 GeV) & 1456 & 68 & 63 & 62 & 62 & 61 \\
   SIMP (400 GeV) & 154 & 14 & 13 & 13 & 13 & 13 \\
   SIMP (700 GeV) & 7 & 1 & 1 & 1 & 1 & 1 \\
   SIMP (1000 GeV) & 1 & 0 & 0 & 0 & 0 & 0 \\
   \hline
 \end{tabular}
 \caption{\label{tab:results_0.1} 
	  The number of events after the selection cuts and several cuts on the charged energy fraction, for the 2 background samples and the 7 different signal samples with $\sigma_{\chi N} / \sigma_{\rm QCD} = 0.1$.}
\end{table}

The results show readily that the considered SIMP signal can be discovered at the LHC for low SIMP mass. At high mass, the sensitivity fades out as the signal production cross section drops. We calculate the signal significance as a function of SIMP mass using the Z-value given by~\cite{Cousins:2008zz}
\be
Z = \sqrt{2}\mathrm{erf}^{-1}(1-2p)
\ee
with p-value
\be
p = B\left(\frac{1}{1 + \tau}, s+b, \tau b + 1\right)
\ee
where $B$ is the incomplete beta function defined in~\cite{bib:incomplete_beta}, $s$ is the number of signal events, $b$ is the number of background events, and $\tau$ is the number of background events divided by the background uncertainty squared.
For this study, we do not provide estimates of the expected precision on the background prediction, but rather consider three levels of uncertainty, 20\%, 50\%, and 100\%, as such providing a range in which the impact of the background uncertainties is demonstrated.
In Tables~\ref{tab:Z_1} and~\ref{tab:Z_0.1}, the Z-value is given for the considered background uncertainties and several CHEF cuts, respectively for the SIMP samples with $\sigma_{\chi N} / \sigma_{\rm QCD} = 1$ and $\sigma_{\chi N} / \sigma_{\rm QCD} = 0.1$.

\begin{table}[ht]
\centering
\renewcommand{\arraystretch}{1.2}
\renewcommand{\tabcolsep}{3.5pt}
\begin{tabular}{| c | c | c | c | c || c | c | c | c || c | c | c | c |}
 \hline
 \multirow{2}{*}{Sample} & \multicolumn{4}{c||}{20\%} & \multicolumn{4}{c||}{50\%} & \multicolumn{4}{c|}{100\%} \\
 \cline{2-13}
  & 0.1 & 0.05 & 0.04 & 0.03 & 0.1 & 0.05 & 0.04 & 0.03 & 0.1 & 0.05 & 0.04 & 0.03 \\
  \hline
  SIMP (1 GeV) & 26 & $\gg$ & $\gg$ & $\gg$ & 10 & 32 & $\gg$ & $\gg$ & 4.9 & 16 & 24 & 34 \\
  SIMP (10 GeV) & 26 & $\gg$ & $\gg$ & $\gg$ & 10 & 32 & $\gg$ & $\gg$ & 4.9 & 16 & 24 & 34 \\
  SIMP (100 GeV) & 23 & $\gg$ & $\gg$ & $\gg$ & 9.0 & 29 & $\gg$ & $\gg$ & 4.3 & 14 & 21 & 30 \\
  SIMP (200 GeV) & 16 & $\gg$ & $\gg$ & $\gg$ & 6.4 & 21 & 31 & $\gg$ & 2.9 & 11 & 16 & 23 \\
  SIMP (400 GeV) & 5.0 & 20 & 28 & 36 & 1.8 & 8.4 & 13 & 18 & 0.52 & 4.0 & 6.3 & 9.4 \\
  SIMP (700 GeV) & 0.33 & 2.7 & 4.6 & 6.7 & $\ll$ & 1.0 & 2.0 & 3.4 & $\ll$ & 0.11 & 0.67 & 1.5 \\
  SIMP (1000 GeV) & $\ll$ & 0.13 & 0.34 & 0.64 & $\ll$ & $\ll$ & $\ll$ & 0.14 & $\ll$ & $\ll$ & $\ll$ & $\ll$ \\
   \hline
 \end{tabular}
 \caption{\label{tab:Z_1} 
	  The Z-values of the signal samples with $\sigma_{\chi N} / \sigma_{\rm QCD} = 1$ for a background uncertainty of 20\%, 50\% or 100\% and CHEF cuts of 10\%, 5\%, 4\% or 3\%. The algorithm used to calculate these significances breaks down at very high or very low values, due to numerical imprecisions. This is represented by $\gg$ and $\ll$ respectively.}
\end{table}

\begin{table}[ht]
\centering
\renewcommand{\arraystretch}{1.2}
\renewcommand{\tabcolsep}{3.5pt}
\begin{tabular}{| c | c | c | c | c || c | c | c | c || c | c | c | c |}
 \hline
 \multirow{2}{*}{Sample} & \multicolumn{4}{c||}{20\%} & \multicolumn{4}{c||}{50\%} & \multicolumn{4}{c|}{100\%} \\
 \cline{2-13}
  & 0.1 & 0.05 & 0.04 & 0.03 & 0.1 & 0.05 & 0.04 & 0.03 & 0.1 & 0.05 & 0.04 & 0.03 \\
  \hline
  SIMP (1 GeV) & 1.0 & 6.1 & 9.4 & 13 & 0.16 & 2.5 & 4.3 & 6.7 & $\ll$ & 0.92 & 1.9 & 3.3 \\
  SIMP (10 GeV) & 1.0 & 6.1 & 9.4 & 13 & 0.15 & 2.5 & 4.3 & 6.6 & $\ll$ & 0.91 & 1.9 & 3.2 \\
  SIMP (100 GeV) & 0.81 & 5.1 & 8.0 & 11 & 0.068 & 2.1 & 3.7 & 5.8 & $\ll$ & 0.68 & 1.6 & 2.8 \\
  SIMP (200 GeV) & 0.44 & 3.3 & 5.4 & 7.9 & $\ll$ & 1.3 & 2.4 & 4.0 & $\ll$ & 0.24 & 0.89 & 1.8 \\
  SIMP (400 GeV) & $\ll$ & 0.73 & 1.4 & 2.2 & $\ll$ & 0.077 & 0.45 & 1.0 & $\ll$ & $\ll$ & $\ll$ & 0.16 \\
  SIMP (700 GeV) & $\ll$ & $\ll$ & $\ll$ & 0.057 & $\ll$ & $\ll$ & $\ll$ & $\ll$ & $\ll$ & $\ll$ & $\ll$ & $\ll$ \\
  SIMP (1000 GeV) & $\ll$ & $\ll$ & $\ll$ & $\ll$ & $\ll$ & $\ll$ & $\ll$ & $\ll$ & $\ll$ & $\ll$ & $\ll$ & $\ll$ \\
   \hline
 \end{tabular}
 \caption{\label{tab:Z_0.1} 
	  The Z-values of the signal samples with $\sigma_{\chi N} / \sigma_{\rm QCD} = 0.1$ for a background uncertainty of 20\%, 50\% or 100\% and CHEF cuts of 10\%, 5\%, 4\% or 3\%. The algorithm used to calculate these significances breaks down at very high or very low values, due to numerical imprecisions. This is represented by $\gg$ and $\ll$ respectively.}
\end{table}

We can conclude from Table~\ref{tab:Z_1} that a discovery can be made through the observation of a 5$\sigma$ excess for SIMP masses up to 400~GeV, for a CHEF cut of 4\% or tighter, for all considered background uncertainties, and for QCD-level SIMP interaction cross section $\sigma_{\chi N}$. This result for a scalar mediator holds as well in the case of a vector mediator, for which a small cross section increase needs to be accounted for, as reported in Table~\ref{tab:xsec}.

For the $\sigma_{\chi N}/\sigma_{QCD} = 0.1$ case, with a background uncertainty of 20\% or 50\%, a CHEF cut of 3\% is needed in order to reach discovery up to $m_\chi = 100 \, \mathrm{GeV}$. If the background uncertainty amounts to 100\%, however, no discovery is possible. This observed impact of a large background uncertainty serves as an accuracy benchmark for background prediction methods in eventual data analysis.

\section{On comparison with other constraints}
\label{sec:comp}

Given the possibility of observing trackless jets at the LHC, it is tempting to try and set limits on the interactions of SIMPs and, possibly, to compare with other constraints. For this, we need to know  1) the high energy production cross section of SIMP pairs, as studied here, 2) the high energy inelastic cross of a SIMP with nucleus,  $\sigma_{\chi N}^{IE}$, which is relevant 
for the response of the calorimeters (and some astrophysical constraints, like that from cosmic rays), and 3) the corresponding low energy elastic cross section, as reported in direct detection searches of dark matter, $\sigma_{\chi N}^E$ (and most of the astrophysical constraints). Given a simplified SIMP model (\ref{eq:model}), we may in principle consider calculating these cross sections. This is provided the simplified model holds over a very broad range of energies. Even if this is satisfied, we have to face the fact that the SIMP interactions involve large couplings, so that the validity of perturbative calculations is questionable. For those reasons, we consider the results of the present section as tentative. 
Keeping this in mind, the main point we would like to explore is the following. Given that the SIMP  is required to produce showers in the hadronic calorimeters, what is the low energy elastic cross section on nucleons? In absence of signal, this requirement sets a constraint on the SIMP cross section that  may be compared with other searches.

We begin with the problem of detection of the SIMP. The hadronic showers in the calorimeters will develop through inelastic scattering of the SIMPs with hadrons. As discussed in the previous section, our target value is a few mb, characteristic of standard hadronic cross sections. For instance the proton-proton total cross section at high energies (target frame) is about $40$~mb, and roughly constant for proton momenta $p_{\rm LAB} \gtrsim $ a few GeV \cite{Agashe:2014kda}, while other hadrons have  similar cross sections, in the range $10-40$~mb (corresponding to a geometrical cross section with a radius of about 1 fermi), and similar energy dependence, exhibiting a plateau at high energies, $p_{\rm LAB} \gg $ GeV, which is the regime of deep inelastic scattering (DIS). A common feature is that, at these energies, the DIS cross section is always larger than the elastic one, which decreases with energy. At lower energies, the presence of resonances makes the situation much more complex and moreover dependent on the nature of the projectiles involved in the collisions, including neutrinos \cite{Formaggio:2013kya} or, as a matter of fact, a SIMP. We will simply assume that the overall scale of inelastic scattering may be estimated from DIS; this should be conservative as the existence of  resonances would increase the total cross section. 
In the case of a scalar mediator, we get from (\ref{eq:model}) that the differential DIS cross section is given by
\be
\df{d\sigma_{\chi N}}{dxdy} = \df{g^2_\chi g^2_q}{2\pi} \df{m_N E}{(Q^2+m^2_\phi)^2}\left[\sum_{i=u,d...} y^2 x f_i(x,Q^2) + y^2x \bar f_i(x,Q^2)\right]\qquad \mbox{\rm (scalar mediator)}
\ee
where the functions $f_i ~(\bar f_i)$ are the parton distribution functions (PDFs) of the quarks (resp. antiquarks) inside the proton and $x~=~{Q^2}/{2m_N\nu}$, with $\nu= {E-E'}$, and $y={\nu}/{E}$  the usual Bjorken variables, where $E$ and $E'$ are the energies  of the incoming and scattered SIMPs in the lab frame. Similarly, for the vector mediator we have
\be
\df{d\sigma_{\chi N}}{dxdy} = \df{\tilde g^2_\chi \tilde g^2_q}{4\pi} \df{m_N E}{(Q^2+m^2_\phi)^2}\left[\sum_{i=u,d...} x f_i(x,Q^2) + x \bar f_i(x,Q^2)(1-y)^2\right]\; \mbox{\rm (vector mediator)}
\ee
The behaviour of the total inelastic cross sections as a function of incoming energy $E$ is similar for both mediators. They increase linearly with $E$ for $E \lesssim m_{\phi}^2/ 2 m_N$, and have  a mild dependence on $E$ for higher energies, being almost constant for the range of interest. This may be understood as follows. For high energies, large energy transfer is suppressed by the propagator, so that $Q^2 =2 m_N x \nu\sim m_\phi^2$, at which point the dominant contribution to the cross section comes from small $x$, $x \sim m_\phi^2/2 m_N E$, with $\nu \sim E$.  A similar  behaviour is predicted for ultra high energy neutrino DIS, in which case  $\sigma_{\nu N}^{DIS} \sim E^{0.4}$ for $E \gtrsim m_{Z}^2/ 2 m_N$~\cite{Gandhi:1998ri}.  A major difference with neutrino DIS, is that for  SIMPs the characteristic  $Q^2 \sim m_\phi^2$ may enter the non-perturbative regime if the mediator is very light, $m_\phi \lesssim$ GeV: this is the main reason why we consider  $m_\phi = 1$ GeV. To the extent that the PDF may be trusted at small $x$ and $Q^2$,\footnote{We have used the MSTW 2008 PDFs in this section \cite{Martin:2009iq}.} from the expressions of the DIS cross sections, we  extract the couplings required for a given inelastic cross section. Taking for simplicity the couplings of the mediator to the light quarks and the DM  to be equal, $g_\chi = g_q \equiv g$ and requiring that $\sigma_{\chi N}^{DIS} = 10$~mb for reference,  we need $\tilde g = 3$ for $m_\phi = 1$ GeV for the vector mediator and  $g = 5.5$ for $m_\phi = 1$ GeV  for the scalar mediator. We notice that, in the case of the scalar mediator, $ g^2/4 \pi \gg 1$, while they  are still perturbative $\tilde g^2/4 \pi \lesssim 1$ for the vector mediator. The difference between the two cases stems from the different couplings to the PDFs. 


From this choice of parameters, we may now estimate the corresponding low energy elastic cross sections.   From (\ref{eq:model}), the SIMP-nucleon elastic cross section through  either the vector or the scalar mediator is given (in the Born approximation--see below) by
\be
\sigma_{\chi N} 
\simeq \df{{g}^2_\chi g^2_q}{\pi m^4_\phi} \mu^2_{\chi N} f_N^2
\ee
where $\mu_{\chi N}$ is the SIMP-nucleon reduced mass and $f_N$ is the effective coupling to the nucleons. 
For the scalar mediator, we have
\be
f_N = m_N \left\{\sum_{q=u,d,s}\df{f^N_{Tq}}{m_q} + \df{2}{27} f^N_{TG}\sum_{q=c,b,t}\df{1}{m_q}\right\}
\label{fN}
\ee
 where $f^N_{Tq}$, defined as $m_N f_{Tq}^N=  \langle N\vert m_q \bar q q\vert N\rangle$, represents the contributions of the light quarks to the nucleon, and $f^N_{TG} = 1 - \sum_{u,d,s} f_{Tq}^N$ \cite{Shifman:1979eb}. Using the values given in \cite{Cerdeno:2010jj}, we get $f_N \approx 11$ (this is summing over all the quarks; taking only the first generation of quarks gives instead $f_N \approx 9.8$). From the coupling required for inelastic scattering we get $g_{\chi N} \approx 55$, which is much larger than the acceptable value, $g_{\chi N} \lesssim 6$, see section \ref{sec:mod}. Hence the scalar mediator scenario is not viable, at least in the simplified framework we consider. In the case of  the vector mediator  we have instead $f_N =3$, which corresponds to $g_{\chi N} \approx 9$. This is about 50 $\%$ larger than what we required, but the difference can be absorbed by adjusting the relative value of the DM and quarks couplings. 

\begin{figure}[hbt]
\centering
\includegraphics[width=0.49\textwidth,angle=0]{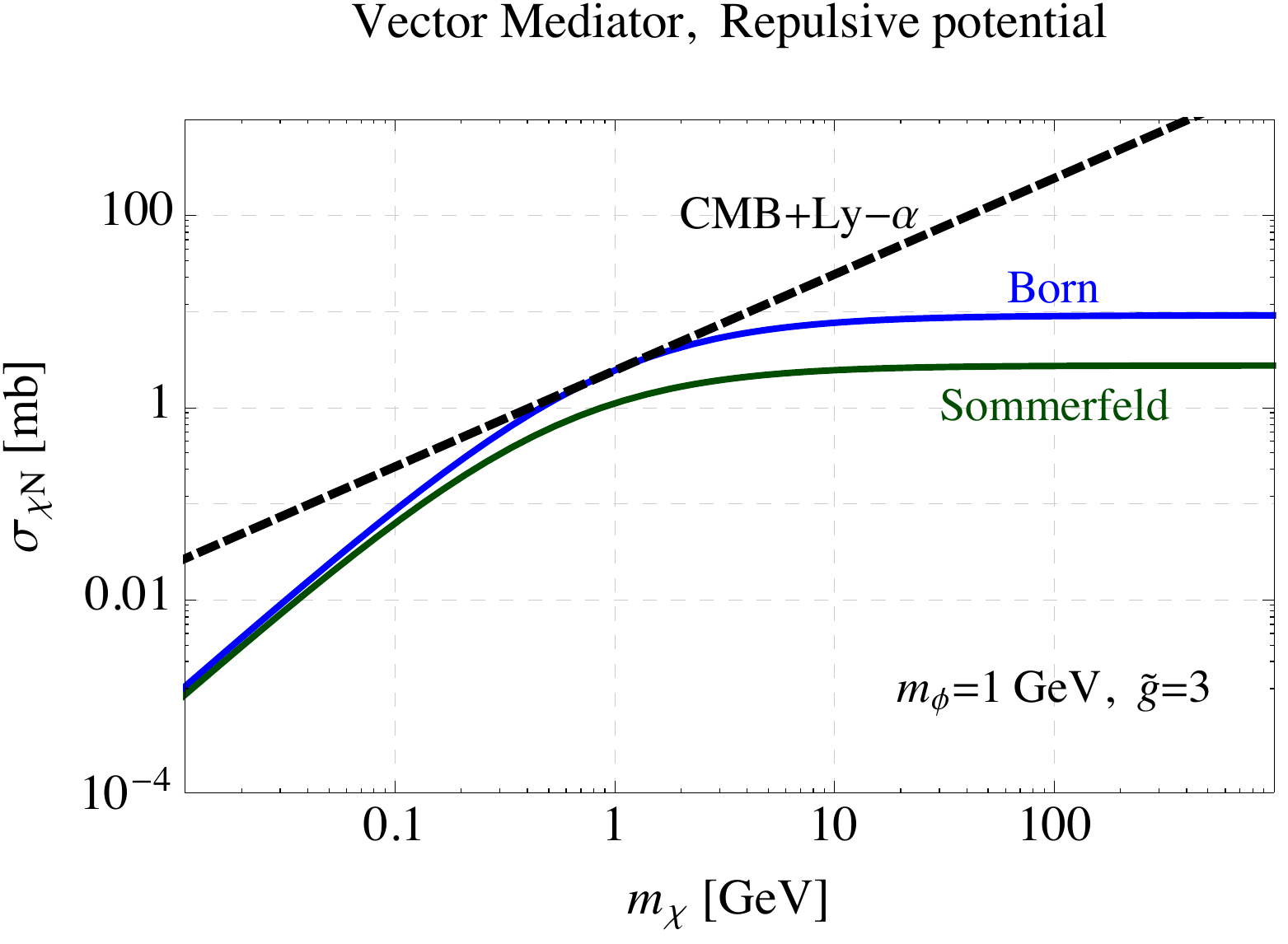}
\includegraphics[width=0.49\textwidth,angle=0]{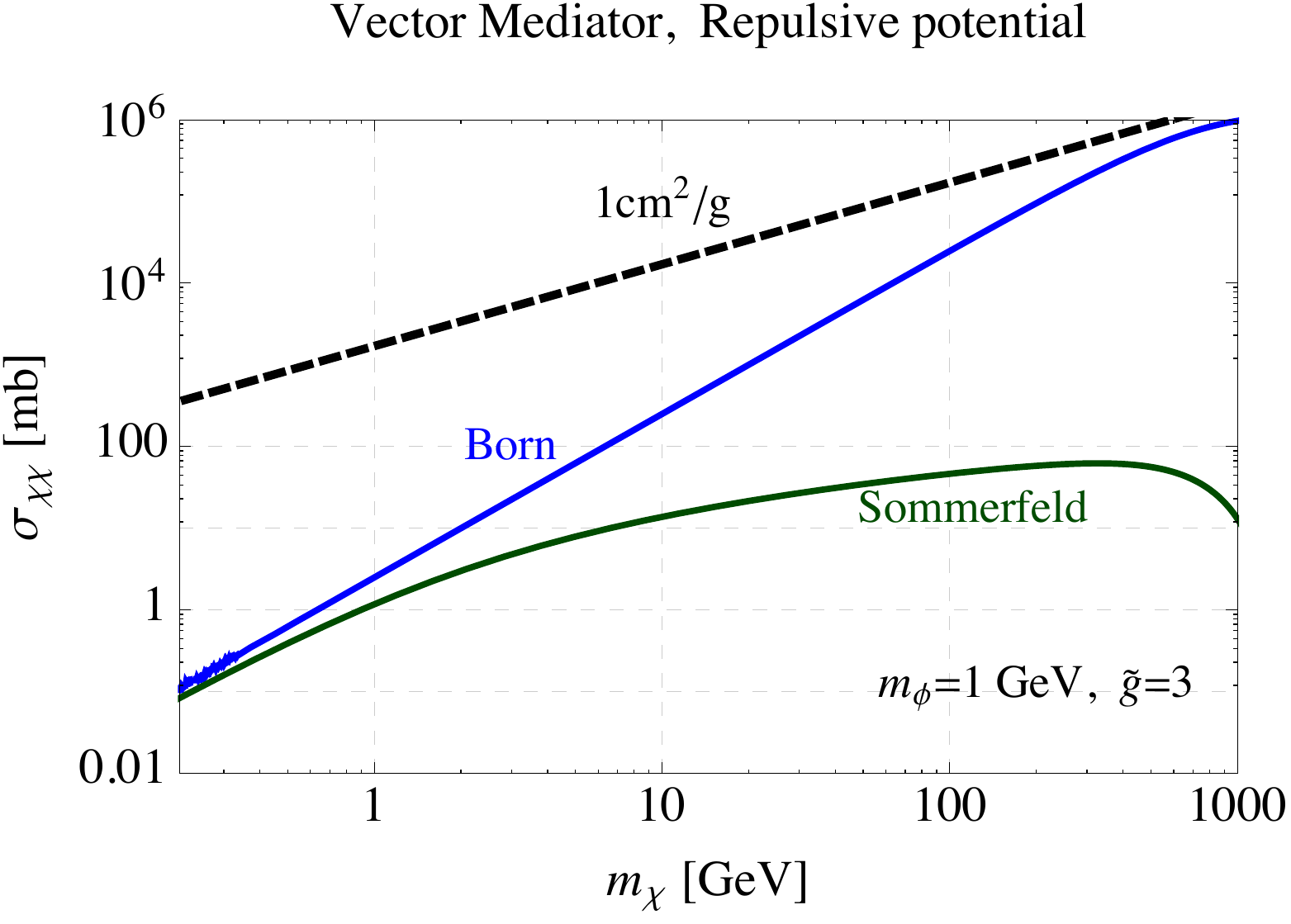}
\caption{\footnotesize Left) Elastic $\chi N$ scattering for the vector mediator case with $m_{\phi}=1$ GeV with couplings $g=3$. The perturbative (Born) approximation (blue line) 
is compared to non-perturbative result (green line). The short dashed line is the constraint from CMB+Lyman-$\alpha$ of Ref.\cite{Dvorkin:2013cea}. Right) SIMP self-interaction for the same benchmark model of the left panel, comparing as well Born with the non-perturbative result. Colour code is identical to the left panel. The short-dashed line corresponds to an astrophysical reference limit on DM self-interactions $\sigma_{\chi \chi}/m_\chi = 1 $cm$^2/$g \cite{Randall:2007ph,Peter:2012jh}. Notice that the astrophysical constraint on interaction with baryons is more stringent than that on the self-interaction, as one would expect.  
}
\label{fig:xsec}
\end{figure}

This being said, we may question whether the Born approximation we used to calculate the low energy elastic cross section is valid for such large couplings. To check this, we have solved the Schr\"odinger equation for $\chi N$ and $\chi \chi$ scattering in the non-relativistic limit. This is akin to calculations of the so-called Sommerfeld effect, but here applied to repulsive interactions, see {\em e.g.} \cite{McDermott:2011jp}.  Interestingly, the Sommerfeld effect is not only non-negligible but also helps in bringing the candidate in agreement with the astrophysical constraints on SIMP interaction with ordinary matter. The strongest limits have been reported in \cite{Dvorkin:2013cea}, based on the impact of DM-baryon interactions on the matter power spectrum, which may be constrained using CMB anisotropies and Lyman-$\alpha$ measurements (see also \cite{Chen:2002yh}). Our results for the elastic cross sections, $\sigma_{\chi N}$ and $\sigma_{\chi\chi}$, are shown in Figure~\ref{fig:xsec}. The figure on the left shows the upper limit on $\sigma_{\chi N}$ from \cite{Dvorkin:2013cea} together with  the low energy cross section we have obtained solving the Schr\"odinger equation (the Born-approximation solution is shown for reference). The figure on the right gives the corresponding $\sigma_{\chi\chi}$ self-interaction cross section\footnote{More rigorously, the so-called ``transverse'' cross section, as defined in \cite{Tulin:2013teo}. }, together with the typical constraint from astrophysical observations, here from the Bullet cluster \cite{Randall:2007ph}. It is important to appreciate that the cross sections are systematically smaller than their Born approximation value because our SIMP candidate has repulsive interactions.\footnote{For self-interactions, the Born approximation breaks down if the particles are non-relativistic and $\tilde g^2/4\pi \times m_\chi/m_\phi \gtrsim 1$ \cite{McDermott:2011jp}. For scattering with nucleons, the dependence on $m_\chi$ is replaced by the SIMP/nucleon reduced mass, $m_\chi \rightarrow \mu_{\chi N}$, which explains why the non-perturbative result becomes independent of the SIMP mass for $m_\chi \gg 1$ GeV).} Notice that in both  figures we implicitly assume that the SIMP is the dominant form of DM.

\begin{figure}[hbt]
\centering
\includegraphics[width=0.7\textwidth,angle=0]{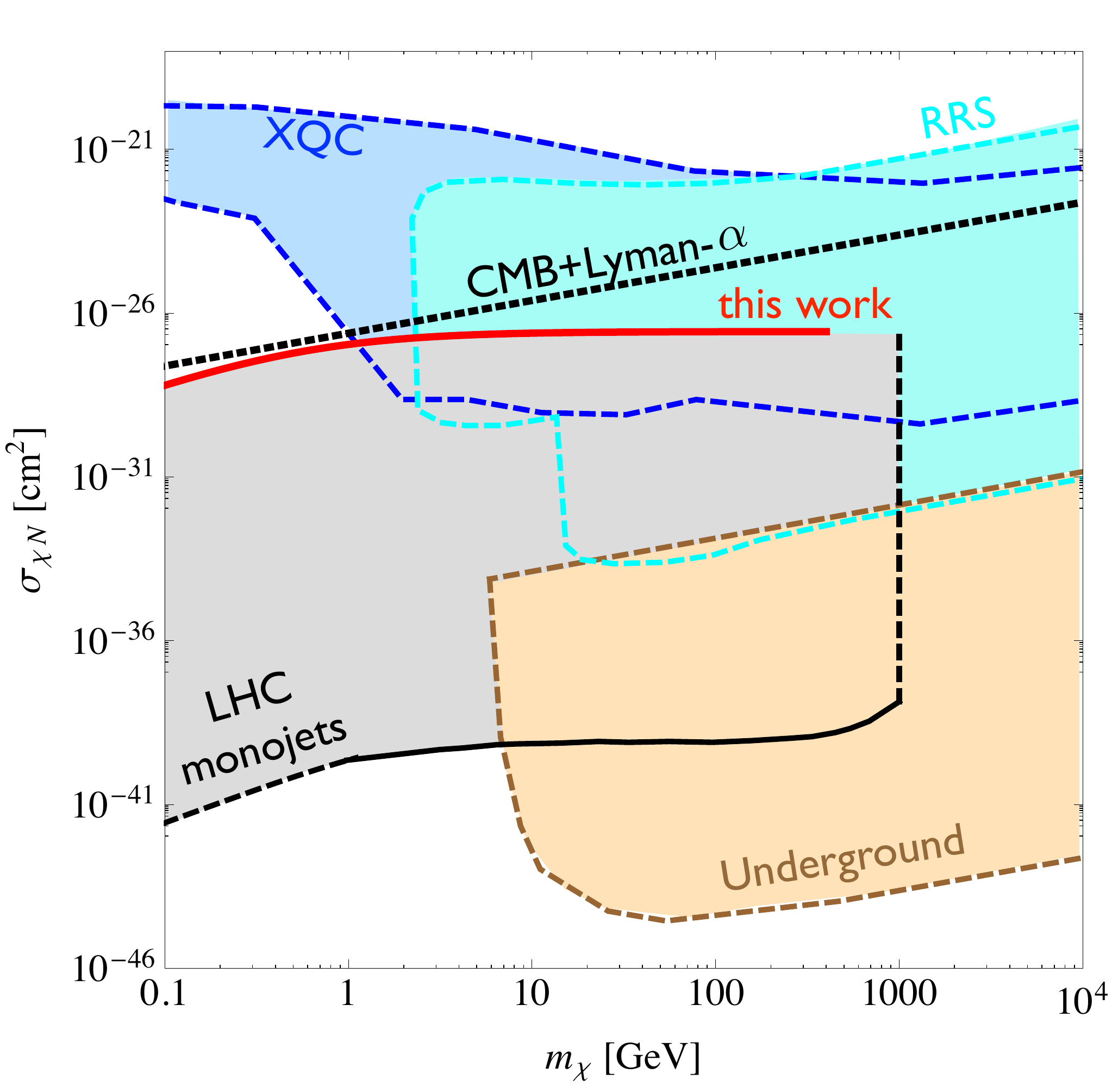}
\caption{\footnotesize Summary plot showing all the most important applicable constraints. Our results are shown in the upper solid red line (``this work''), which corresponds to the green line of Figure~\ref{fig:xsec} (left). In black solid/dashed (lower lines), the monojet constraints are shown. The other constraints are: atmospheric XQC and RRS experiments (blue and cyan, respectively), underground experiments (brown dashed), and CMB+Lyman-$\alpha$ (black dashed). 
}
\label{fig:allcons}
\end{figure}

We now proceed to bring everything together. In Figure~\ref{fig:allcons}, we show the relevant existing constraints in the $\sigma_{\chi N}-m_{\chi}$ plane. In brief, the figure shows the exclusion limits from direct detection searches, which are inoperant for large cross sections \cite{Albuquerque:2003ei}, those from various astrophysical observations that we just discussed  and from rocket or balloon experiments (see Section~\ref{sec:mod}), and finally those from DM searches at colliders. Most relevant for the present discussion is the line that corresponds to missing energy searches. The constraints are very strong for low mass DM candidates, complementary to those from direct searches. As the cross section of DM with nucleons increases however, at some point the signature is no longer missing energy, but trackless jets. This corresponds to the thick solid (red) line around $\sigma_{\chi N} \approx 10^{-28}$ cm$^2$ for $m_\chi \gtrsim 1$ GeV. If no excess above the QCD background is measured at the LHC, then the parameter space above this line would be excluded. The limit is  conservative but  robust,  as a larger coupling implies a larger number of events (we took $g=1$ in Section~\ref{sec:res}), while the response of the detector would drop rapidly for a smaller coupling. We have tentatively extended the limit all the way up to $m_\chi = 400$ GeV, consistent with the analysis reported in Table~\ref{tab:Z_1} and~\ref{tab:Z_0.1}, while we are aware that our discussion of deep inelastic scattering is probably not completely reliable for such heavy SIMP candidates. As for the collider searches for WIMPs, the search of SIMPs is complementary with other constraints (essentially astrophysical in nature) for light candidates, below say 1 GeV. As is usual for such considerations, we emphasize again that the SIMP particles produced may not be the dominant form of DM, in which case the astrophysical constraints are essentially irrelevant.

\section{ \label{sec:con} Conclusion}

In this work we have considered further the possibility that DM may be made (partially or totally) of particles with strong interactions with ordinary matter. These so-called SIMPs, for strongly interacting massive particles, are much less considered than their more popular siblings, the WIMPs, but they are regularly considered in the literature in order to address some astrophysical issues. While they are  challenged by many observations, again mostly astrophysical, they are not completely excluded. Furthermore, little work has been done on possible constraints from colliders. Extending on previous works, in particular \cite{Bai:2011wy}, we have studied in more details the possibility of observing trackless jets at the LHC, taking into account realistic simulations of the QCD background and the response of the detectors. Most notably, we show that the charged content of jets is a powerful discriminator to suppress dijet backgrounds at LHC, thus enhancing the sensitivity to a potential SIMP signal. Our analysis shows that SIMPs with mass up to $m_\chi \sim 400$ GeV could lead to an observable signal, provided its interaction cross section with ordinary matter is about 10$\%$ of that of ordinary nucleons. Most of our work is dedicated to the forecast for the experimental search of SIMPs at the LHC. To do so, we have adopted a simplified SIMP model. In this framework the SIMP interacts with quarks through a light mediator particle. This framework also allows, at least in principle, to compare the sensitivity reach of the LHC search with other constraints, most of which are at much lower energies. We show this in Figure~\ref{fig:allcons}, assuming that our SIMP simplified candidate constitutes the dominant form of dark matter. Where relevant, we pointed out the limitations due to the difficulty of doing reliable calculations with strongly interacting particles, be them within a simplified framework. The signature and the proposed analysis are however essentially independent of these potential complications, {\em i.e.} they stand by themselves. While much improvement could also be envisioned for the experimental analysis, most notably a more realistic analysis of the detector response, our feasibility study shows that SIMP candidates over a wide mass range may be efficiently searched at the LHC. 
 
\section
*{Acknowledgements}
The authors want to thank M. Fairbairn, K. Mawatari, P. Schwaller and P. Vanlaer for very fruitful discussions.  The work of BZ and MT is supported by the IISN and by the Belgian Federal Science Policy through the Interuniversity Attraction Pole P7/37. SL, ND, and IDB are supported through FWO Odysseus II grant G.0C39.13N.



\bibliography{bibliography}

\end{document}